\documentclass[11pt,a4paper]{article}
\usepackage{jheppub}
 \usepackage{etoolbox}
\makeatletter
\patchcmd{\maketitle}{\@fpheader}{}{}{}
\makeatother

\usepackage[english]{babel}
\usepackage[T1]{fontenc}
\usepackage[utf8]{inputenc}

\usepackage{bm}

\begin{document}
\newcommand{\eg}{{\it e.g.}}
\newcommand{\etal}{{\it et. al.}}
\newcommand{\ie}{{\it i.e.}}
\newcommand{\be}{\begin{equation}}
\newcommand{\dd}{\displaystyle}
\newcommand{\ee}{\end{equation}}
\newcommand{\bea}{\begin{eqnarray}}
\newcommand{\eea}{\end{eqnarray}}
\newcommand{\bef}{\begin{figure}}
\newcommand{\eef}{\end{figure}}
\newcommand{\bce}{\begin{center}}
\newcommand{\ece}{\end{center}}

\title{Thermodynamic Geometry and Deconfinement Temperature}
\author[a,b]{P. Castorina}
\author[a]{M. Imbrosciano}
\author[a,b]{Lanteri}
\affiliation[a]{Dipartimento di Fisica, Universit\`a di Catania, Via Santa Sofia 64,
	I-95123 Catania, Italy.}
\affiliation[b]{INFN, Sezione di Catania, I-95123 Catania, Italy.}

\abstract{
The application of Riemannian geometry to the analysis of the equilibrium thermodynamics in Quantum Chromodynamics (QCD) at finite temperature and baryon density gives a new method to evaluate the critical temperature, $T_c$, of the deconfinement transition. In the confined phase, described by the thermodynamic geometry of the Hadron Resonance Gas, the estimate of $T_c$ turns out completely consistent with  lattice QCD simulations of the quark-gluon plasma phase  if the hadron excluded volume and the interaction effects are taken into account.  }


\maketitle

\section{Introduction}

The application of Riemannian geometry to thermodynamic systems \cite{Rao,Weinhold,Rupp1979} is based on the introduction, through the Hessian of the
entropy density, of a metric tensor, i.e. a measure of  ``distance'', in phase space.

The main results of thermodynamic geometry (TG) are the  (inverse) relation between the line element 
and the fluctuation probability between equilibrium states and the, so called, ``interaction hypothesis'', which states that the absolute value of the scalar curvature $R$, an intensive variable (with units of a volume) evaluated by the metric, is proportional to  $\xi^3$, where $\xi$ is the correlation length of the thermodynamic system. Indeed, a covariant and consistent thermodynamic fluctuation theory can be developed~\cite{Ruppeiner:1995zz}, which  generalizes the classical one and offers a theoretical justification to the physical meaning of $R$. 

TG  has been tested in many different systems: phase coexistence for Helium, Hydrogen, Neon and Argon~\cite{Ruppeiner:2011gm}, for the Lennard-Jones fluids~\cite{Ruppeiner2013,May2012}, for ferromagnetic systems and liquid-liquid phase transitions~\cite{Dey:2011cs}. Another field of application  is the study of the phase transitions of cosmological interest: the liquid-gas like first order phase transition in dyonic charged AdS black hole~\cite{Chaturvedi:2014vpa} and in the Hawking-Page transitions in Gauss-Bonnet-AdS black holes~\cite{Sahay:2017hlq}. 

Very recently \cite{CIL} the thermodynamic geometry method has been applied, for the first time, to Quantum-Chromodynamics (QCD), at large temperature and low baryon density, to evaluate the (pseudo-)~critical deconfinement temperature $T_c$. Lattice QCD data~\cite{Bazavov:2017dus}  and the Hadron Resonance Gas model (HRG) have been used to evaluate the thermodynamic potentials respectively in the quark-gluon plasma and in the confined phases  and, since the transition is a cross-over, $T_c$ has been determined by the criterion $R=0$ (see ref.~\cite{CIL} for details).

The deconfinement temperature evaluated by HRG turns out to be  larger than the Lattice QCD result. This difference could be of physical origin, suggesting the interpretation that the meson melting temperature is larger than the  temperature associated with the chiral susceptibility, or it could be an artifact of the specific HRG model, since point-like hadrons have been considered in ref.~\cite{CIL}. 

In fact, the introduction of other dynamical details, as the excluded volume, could change the HRG evaluation of $R$, by including some effective repulsive interaction similar to Fermi statistic effects and then closing, in part, the gap with the value of $R$ for a mostly fermionic system as the quark-gluon plasma. 

This analysis is carried out in this paper.  In Sec.~\ref{sec:1} the thermodynamic geometry approach is briefly recalled.
 Sec.~\ref{sec:2} is devoted to the different HRG models and the results are given in Sec.~\ref{sec:3}.
The ``discrepancy'' between Lattice QCD estimate of $T_c$ and the HRG one disappears by including the excluded volume and interaction effects.

\section{Thermodynamic Geometry and R=0 criterion \label{sec:1}}

Let us consider an open thermodynamic system, of fixed volume $V$ and  $r$ species of constituents, described in terms of the standard densities $a^i = (a^0, a^1 \dots a^r)$, where $a^0$ is the internal energy density and the other components are the number of particles of the different species.

Calling $s(a)$ the entropy density of the system, the fluctuation probability density, $P_f$, in Gaussian approximation is given by \cite{Ruppeiner:1995zz}
\be P_f \propto exp \left\lbrace - \frac{V}{2}g_{ij} \Delta a^i \Delta a^j \right\rbrace, \ee
where 
\be
\label{g}
g_{ij} = - \frac{1}{k_B} \frac{\partial ^2 s(a)}{\partial a^i \partial a^j}
\ee
is the metric tensor and the line element $(\Delta l)^2 = g_{ij} \Delta a^i \Delta a^j$ is an invariant, positive definite quadratic form (as a consequence of the maximum entropy principle, see~\cite{Ruppeiner:1995zz} for detailed explanations). Starting from Eq.~\eqref{g}, the calculation of the scalar curvature $R$ is straightforward; we will use the standard intensive quantities in the entropy representation:
\be 
F^\mu \equiv \frac{\partial s(a)}{\partial a^\mu} = \left( \frac{1}{T}, -\frac{\mu^1}{T} \dots -\frac{\mu^r}{T} \right),
\ee 
where $\mu^i$ are the chemical potentials of the different species and $T$ is the overall temperature. In this ``frame'', the metric depends on the derivatives of the thermodynamic potential $\phi = P/T$, where $P$ is the total pressure of the system~\cite{Ruppeiner:1995zz}. 

In two dimensions, the scalar curvature $R$ contains all the informations about the geometry of the Riemannian manifold and its expression is considerably simplified:
\be\label{eq:R}
R = \frac{k_B}{2 g^2} 
\begin{vmatrix}
\phi_{,11} & \phi_{,12} & \phi_{,22} \\
\phi_{,111} & \phi_{,112} & \phi_{,122} \\
\phi_{,112} & \phi_{,122} & \phi_{,222}
\end{vmatrix}\;,
\ee
where $k_B$ is the Boltzmann's constant, $g$ is the determinant of the metric and the usual comma notation for derivatives has been used. For example,  $\phi_{,12}$ indicates the derivative of $\phi$ with respect to the first coordinate $\beta=1/T$ and the second coordinate ($\gamma=-\mu/T$).  The metric elements are
\begin{equation}
g_{ij} = \frac{1}{k_B}\;\phi_{,ij} 
\end{equation}

As discussed in the introduction, the interaction hypothesis has been confirmed for a relevant number of different systems.
In particular, the most interesting developments are in the field of real fluids where the absolute value of $R$ is a direct measure of the size of organized mesoscopic fluctuating structures in thermodynamic systems~\cite{Rupp2017}.

The criterion $R=0$ has been applied~\cite{CIL} to  QCD  by considering two thermodynamic variables, $\beta=1/T$ and $\gamma^2$, i.e.  a 2-dimensional thermodynamic metric. 
The potential $\phi=P/T$ is evaluated by lattice QCD data in the quark-gluon phase~\cite{Bazavov:2017dus}, available at small chemical potential, and  then the calculations are  reliable in the range $\gamma^2 < 1$, where a series expansion is feasible. However the criterion $R=0$ is completely general.

\begin{figure}
	\centering 
	\includegraphics[width=0.5\columnwidth]{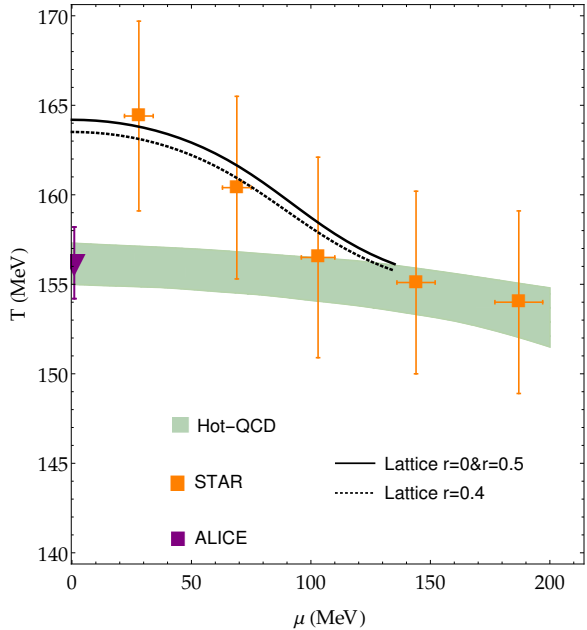}
	\caption{The crossing temperature evaluated by $R=0$, both for $n_Q=n_S=0$ (continuous black line) and for $n_S=0$ and $n_Q/n_B=0.4$ (black dotted line), compared with lattice data (light-blu-gray band) and the results of the freeze out temperature from ALICE (purple point~\cite{Floris:2014pta}) and STAR (orange points~\cite{Das:2014qca,Adamczyk:2017iwn}) collaborations.}
	\label{fig:TR}
\end{figure}

By the expansion of the  potential  $\phi$ as a power series in  $\gamma^2$ around the point $\mu_B=0$ (see Ref.~\cite{CIL} for details),  the (pseudo-)~critical temperature $T_c$, evaluated by the criterion $R=0$ is reported in Fig.~\ref{fig:TR} and compared with lattice results on the critical temperature deduced by chiral susceptibility~\cite{Bazavov:2017dus,qm18} (light-blu-gray band) and the freeze-out temperature  obtained by ALICE~\cite{Floris:2014pta} (purple point) and STAR~\cite{Das:2014qca,Adamczyk:2017iwn} collaborations (orange points). 
The continuous black curve is from lattice data obtained by the condition $n_S=n_Q=0$, while the dotted black one is for $n_S=0$ and $n_Q/n_B =0.4$
($n_S$, $n_Q$, $n_B$ being the strangeness, charge and baryon number densities respectively), following the procedure of Ref.~\cite{Bazavov:2017dus,qm18} where $n_Q$ is considered as a function of $\mu_B$.
The calculation based on the $R=0$ criterion agrees with lattice QCD results within $10\%$.

\section{The criterion $R=0$ for different Hadron Resonance Gas models\label{sec:2}}

For temperature below $T_c$, the confined phase can be described by a gas of hadrons and resonances. There are, however, various models of the HRG, which essentially differ
in the treatment of the excluded volume and/or of the residual interactions.

Independently on the particular HRG model, by the expansion of the pressure as a power series of $\gamma^2=\left(-\mu_B/T\right)^2$ around $\mu_B=0$, i.e.
\begin{equation}\label{eq:P}
P(\beta,\gamma) = P_0+P_2\;\gamma^2 + P_4\;\gamma^4
+P_6\;\gamma^6\cdots \;,
\end{equation}
the scalar curvature $R$ can be written as
\begin{equation}\label{eq:Rs}
R(\beta,\gamma)
=
\sum_{n=0}^\infty
R_{\mathcal O(2\,n)}\;\gamma^{2\,n}
\;,
\end{equation}
where any coefficients $R_{\mathcal O(2n)}$ is function of the first $(2n-1)$ terms of \eqref{eq:P}  (the first terms in~\eqref{eq:Rs} are evaluated in~\cite{CIL}). 

In the ideal (id) HRG model with point-like constituents,  in the Boltzmann approximation, the pressure is given by the sum of all, non interacting, hadron and resonance contributions up to a maximum mass $m_{max}$:
\begin{equation}
p_{HRG}^{id}(T,\mu) = T\; \sum_{m_i\leq m_{max}} 
n_i^{id}(T,\mu)	\;,
\end{equation}
where
\begin{equation}
n_i^{id} =\frac{g_i\;m^2_i\;T}{2\,\pi^2}\; e^{\frac{\mu_i}{T}}\;K_2\left(\frac{m_i}{T} \right)
\end{equation}
is the number density of the $i-$th resonance/particle (with mass $m_i$ and degeneracy $g_i$) and $K_2(x)$ is the Macdonald function (see, for example,~\cite{Vovchenko:2017}). 

Recently, some classical models of equation of state (EoS) for real gases have been extended in order to include quantum statistical effects  and to apply the EoS to hadronic and nuclear systems. 

In a general excluded volume (ev) model, in the Boltzmann approximation, the pressure can be written as~\cite{Vovchenko:2017}
\begin{equation}\label{eq:pev}
p_{ev}= T\,Z(\eta)\,n
\;,
\end{equation}
where $\eta =b\,n/4= 4\,\pi\,n\,r^3/3$  is the packing fraction, $n$ is the  number density and $Z(\eta)$ is a dimensionless ``compressibility'' factor.

Specific models can be obtained by defining, for example,  an excluded volume repulsion term by the van der Waals (VdW) equation as 
\begin{equation}\label{eq:VdWZ}
Z_{VdW}(\eta) = \frac{1}{1-4\;\eta} 
\end{equation}
or by the Carnahan-Starling (CS) term~\cite{Carnahan} 
\begin{equation}\label{eq:fCSZ}
Z_{CS}(\eta) = \frac{1+\eta+\eta^2-\eta^3}{(1-\eta)^3}
\;.
 \end{equation}
The CS compressibility factor~\eqref{eq:fCSZ},  introduced to describe a gas of rigid spheres and to improve  the approximation of the virial expansion~\cite{Carnahan}, indeed reproduces rather accurately the virial expansion terms up to the eighth order, where the VdW approach fails (recall that higher order terms describe the contribution of non-binary interactions).

The previous expression~\eqref{eq:pev} can be extended to the gran canonical ensemble and generalized to include an attractive interaction by performing a ``shift'' in the chemical potential~\cite{Vovchenko}  
\begin{equation}
\mu=\mu^\star - \frac{b}{4}\,f^\prime(\eta)\,p^{id}(T,\mu^\star) + u(n)+n\,u^\prime(n) 
\end{equation}
where $f(\eta)$ gives the permitted volume region,  $u(n)$ is the mean total energy per particle of the attractive interaction and $n=n(T,\mu)$ is the  number density. 

The pressure turns out to be \cite{Vovchenko}   
\begin{equation}
p(T,\mu) = \left[f(\eta) - \eta\,f^{\prime}(\eta)\right]\,p^{id}(T,\mu^\star) + 
n^2\,u^{\prime}(n)\;,
\end{equation}
and entropy, energy density and number density are related to the corresponding quantities of the ideal HRG  by the equations:
\begin{equation}
s(T,\mu) = f(\eta)\,s^{id}(T\,\mu^\star)\;,
\end{equation}
\begin{equation}
\varepsilon(T,\mu) = f(\eta) \,\varepsilon^{id}(T,\mu^\star) + n\;u(n),
\end{equation}
and
\begin{equation}
n(T,\mu)=f(\eta)\;n^{id}(T,\mu^\star)
\;.
\end{equation}
The quantities~\eqref{eq:VdWZ} and \eqref{eq:fCSZ} are now replaced by
\begin{equation}\label{eq:fVdW}
f_{VdW}(\eta) =1-4\;\eta 
\end{equation}  
and
\begin{equation}\label{eq:fCS}
f_{CS}(\eta) = \exp\left\{-\frac{\left(4-3\;\eta\right)\;\eta}{\left(1-\eta\right)^2}
\right\}\;.
\end{equation}
The attractive term  can have different expressions also. For example, a VdW approach leads to 
\begin{equation}\label{eq:uVdW}
u_{VdW}(n)=-a\;n
\end{equation}
and the Clausius form is
\begin{equation}\label{eq:uC}
u_{Cl}(n) = -\frac{a\;n}{1+b\;n} 
\;.
\end{equation}
Finally, the previous models can be easily generalized to a multi-component gas by defining the ideal pressure as a sum over meson, $M$, baryon, $B$ and  anti-baryon $\overline B$ contributions,
\begin{equation}
p^{id}(T,\mu^\star) = 
\sum_{j\in M}p_j^{id}(T,\mu^\star_j)
+
\sum_{j\in B}p_j^{id}(T,\mu^\star_j)
+
\sum_{j\in \overline B}p_j^{id}(T,\mu^\star_j)
\;,
\end{equation}
with  different packing fraction for each species.

\begin{table}[t]
	\vspace{0.5cm} 
	
	\begin{center}
		\begin{tabular}{|c|c|c|}
			\hline
			&$\quad a\;\left(\text{MeV}\;\text{fm}^3\right)\quad $&$\quad b\;\left(\text{fm}^3\right)\quad $
			\\
			\hline	
			\hline
			VdW & $329$&$3.42$\\
			\hline	
			Clausius-CS&$423$&$2.80$\\
			\hline	
		\end{tabular}
	\end{center}
	\caption{The values of the parameter used in Eqs.~\eqref{eq:uVdW}~and~\eqref{eq:uC}~\cite{Vovchenko}.}
	\label{tab:par}
\end{table}

\section{Results \label{sec:3}}

We have considered three different HRG models: 
\begin{itemize}
	\item the ideal HRG model of point-like constituents;
	\item the Clausius-CS-HRG model, where the repulsive excluded volume interaction is give by the Carnahan-Starling term (See Eq.~\eqref{eq:fCS}) and the attractive one by the Clausius form (See Eq.~\eqref{eq:uC});
	\item the VdW-HRG model which describes a gas of hadrons and resonances where, repulsive and attractive, interactions follow the standard VdW equation (Eqs.~\eqref{eq:fVdW} and~\eqref{eq:uVdW}).
\end{itemize}

In all previous models: a)~only baryon-baryon (anti-baryon - anti-baryon)  interactions have been included, neglecting the other ones; b)~mesons are considered point-like; c)~the parameter $a$ and $b$ for all (anti)baryons are assumed to be equal to those of nucleons and are fixed by the nuclear matter properties~\cite{Vovchenko} (see Tab.~\ref{tab:par}); d)~all strange and non-strange hadrons  in the Particle Data Table have been included with the exception of $\sigma$ and
$\kappa$ mesons (see \cite{Vovchenko:2017,Vovchenko:2017b} for details).

\begin{figure}
	\centering
	\includegraphics[width=0.5\columnwidth]{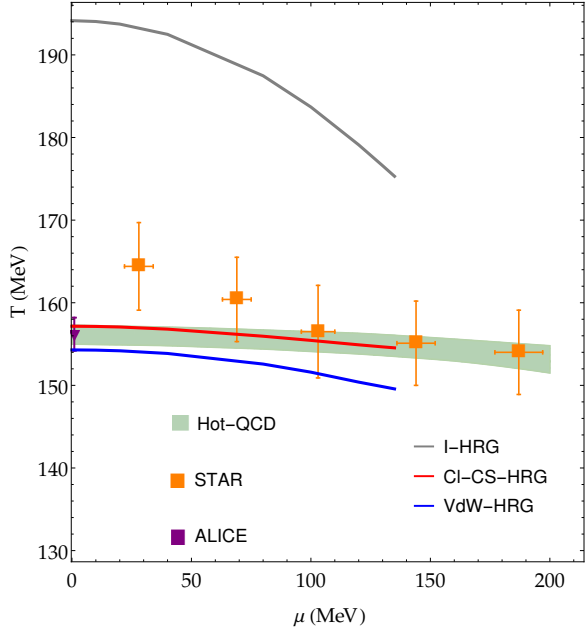}
	\caption{The temperature from the $R=0$-criterion obtained from the ideal-HRG model (gray curve), the Clausius-CS-HRG model (red curve) and the VdW-HRG model (blue curve). 
	The light-blu-gray band is for lattice results on the critical temperature deduced by chiral susceptibility~\cite{Bazavov:2017dus,qm18}, orange points are for the freeze-out temperature  obtained by STAR~\cite{Das:2014qca,Adamczyk:2017iwn} and the purple one by ALICE~\cite{Floris:2014pta} collaborations.}
	\label{fig:TH}
\end{figure}

\begin{figure}
	\centering
	\includegraphics[width=0.5\columnwidth]{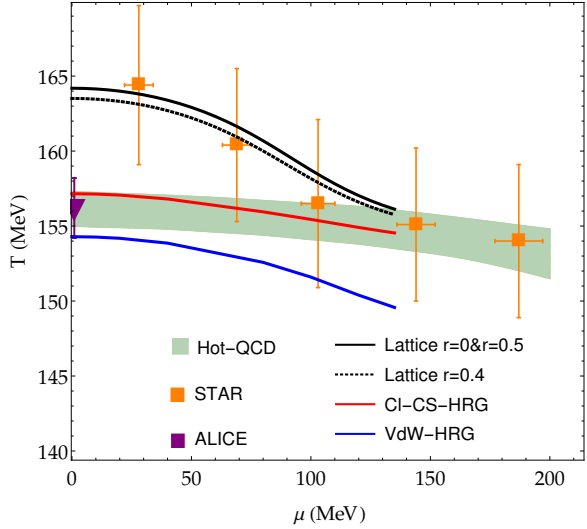}
	\caption{The temperature in the CS-Clausius-HRG model (red curve), in the VdW-HRG model (blue curve) and from lattice data (black curves), compared with lattice results on the critical temperature deduced by chiral susceptibility~\cite{Bazavov:2017dus,qm18} (light-blu-gray band) and the freeze-out temperature  obtained by ALICE~\cite{Floris:2014pta} (purple point) and STAR~\cite{Das:2014qca,Adamczyk:2017iwn} collaborations (orange points).}
	\label{fig:TH2}
\end{figure}

In Figure~\ref{fig:TH} are plotted the temperature from the $R=0$-criterion for the three considered HRG-like models. 
Figure~\ref{fig:TH2} shows the temperature in the CS-Clausius-HRG model (red curve) and in the VdW-HRG (blue curve), compared with that from lattice QCD (black curves).

\section{Comments and Conclusions}

The introduction of a thermodynamic metric and the calculation of the corresponding  scalar curvature, $R$, is a useful tool to estimate the deconfinement temperature. The criterion $R=0$ applied to the quark-gluon plasma phase gives a value of $T_c$ in good agreement with QCD lattice simulations. By the same method one determines the critical line in the confined phase, described by different HRG models, and the result is completely consistent with Lattice QCD data if the hadron excluded volume and the interaction effects are taken into account. Notice the excellent agreement with the CS-Clausius-HRG model.
The approch has been applied for small baryon density since it requires a reliable evaluation of the thermodynamic potential $\phi$. However the geometrical approach  is quite general and  the calculations at large baryon density  can be analytically done once $\phi$ is known.

\end{document}